\documentclass[aps,prl,twocolumn,superscriptaddress]{revtex4}
\usepackage{graphicx}

\newcommand{\omegaS}{\omega_\mathrm{S}}

\newcommand{\omegaONE}{\omega_\mathrm{AOM_1}}
\newcommand{\omegaTWO}{\omega_\mathrm{AOM_2}}

\newcommand{\TE}{\tau_\mathrm{E}}
\newcommand{\ELO}{E_\mathrm{LO}}
\newcommand{\EO}{E}
\newcommand{\EL}{E_\mathrm{L}}
\newcommand{\EI}{E_\mathrm{I}}

\newcommand{\DB}{D_\mathrm{B}}

\begin{document}


\title{Heterodyne holographic microscopy of gold particles.}

\author{Michael Atlan}
\author{Michel Gross}
\author{Pierre Desbiolles}
\affiliation{Laboratoire Kastler-Brossel de l'\'Ecole Normale
Sup\'erieure, CNRS UMR 8552, Universit\'e Pierre et Marie Curie -
Paris 6, 24 rue Lhomond 75231 Paris cedex 05. France}

\author{\'Emilie Absil}
\author{Gilles Tessier}
\affiliation{\'Ecole Sup\'erieure de Physique et de Chimie
Industrielles de la Ville de Paris, CNRS UPR 5, Universit\'e Pierre
et Marie Curie - Paris 6, 10 rue Vauquelin 75231 Paris cedex 05.
France}

\author{Ma\"it\'e Coppey-Moisan}
\affiliation{D\'epartement de Biologie Cellulaire, Institut Jacques
Monod, CNRS UMR 7592, Universit\'e Paris 6 and 7, 2 Place Jussieu,
Tour 43, 75251 Paris Cedex 05. France}

\date{\today}

\begin{abstract}
We report experimental results on heterodyne holographic microscopy
of subwavelength-sized gold particles. The apparatus uses continuous
green laser illumination of the metal beads in a total internal
reflection configuration for dark-field operation. Detection of the
scattered light at the illumination wavelength on a charge-coupled
device array detector enables 3D localization of brownian particles
in water.\\
OCIS : 180.6900, 090.1995, 170.0180
\end{abstract}

\maketitle

\begin{figure}[]
\centering
\includegraphics[width = 8.5 cm]{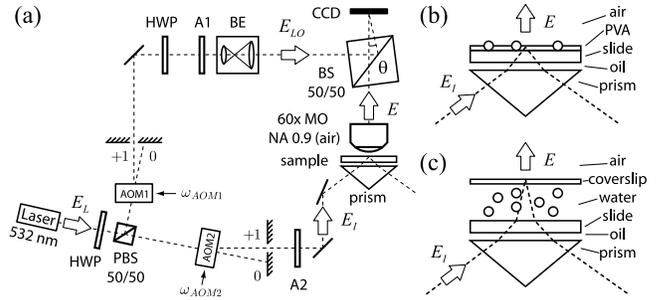}
\caption{Experimental setup (a). Evanescent wave illumination of
static beads in a plane (b). Total internal reflection configuration
for dark-field illumination of beads in a tridimensional environment
(c). Acronyms defined in the text.}\label{fig_setup}
\end{figure}

Assessing microscopic processes by tracking optical labels has a
broad range of applications in biology. In particular,
monitoring biological phenomena such as cellular-level dynamics is
a subject of growing interest. Fluorescent molecules are widely used in this aim, but 
the observation time of dyes
is limited by photobleaching. 
Quantum dots offer a much better photostability, but they have the
inconvenient to blink. 
Noble metal nanoparticles 
have the advantage of being perfectly photostable \cite{Schultz2000}.
Originally, light scattered by small metallic particles
has been detected in  
dark field \cite{Schultz2000} or total internal reflection
\cite{Sonnichsen2000} configuration.  
To improve the detection sensitivity,
interferometric approaches  have been introduced
\cite{BoyerTamarat2002, Lindfors2004, Dijk2005, Ignatovich2006,
Jacobsen2006}.
%
%
%
Common-path interference, initially achieved with a Nomarski
interferometer \cite{BatchelderTaubenblatt1989}, 
translates phase variations 
into intensity variations, and 
enables the detection of
phase perturbation 
provoked by spatial
\cite{Hwang2007} or photothermal \cite{BerciaudLasne2006}
modulation. 
%
%
Scanning heterodyne detection of the photothermal modulation
\cite{BoyerTamarat2002} enables an unmatched combination of
sensitivity and selectivity suitable to discriminate particles
smaller than 5 nm from their background. Photothermal imaging
relies on a spatial scanning of the laser beams to track the index
modulation in the neighborhood of the beads; the temporal noise in
the incident light leads to spatial noise in the image acquired
sequentially \cite{Jacobsen2006}. Wide-field detection schemes
alleviate such issues, but their sensitivity in optical mixing
configurations would not match heterodyne detection on single
detectors levels.
We propose here a 
wide-field, shot-noise limited, tunable CCD heterodyne detection
technique able to achieve high-resolution, 3D microscopy of gold
particles with a laser source at rates compatible with biological
dynamics. The experimental setup is sketched in fig.
\ref{fig_setup}(a).

The main laser beam ($\lambda$ = 532 nm, field $\EL$,
50 mW, single axial mode, CW) 
is split
with a polarizing beam splitter (PBS)
in local oscillator (LO) and illumination arms (fields $\ELO$ and
$\EI$). A $\lambda/2$ waveplate (HWP), and  neutral densities (A1,
A2) allow control of the optical power in each arm. Both beams are
frequency-shifted around 80 MHz by acousto-optical modulators
(AOM) driven 
at frequencies $\omegaONE$, $\omegaTWO$.
The LO beam passes through a beam expander (BE) to form a plane wave
which polarization angle is adjusted with a HWP to maximize the
holographic modulation depth. The object is illuminated in dark
field configuration by using total internal reflection (TIR). The
scattered field $\EO \ll E_{LO}$ passes through a microscope
objective (MO, 60$\times$ magnification, $\rm NA = 0.9$, air).
Off-axis optical mixing ($\theta$ tilt) of $\EO$ with $\ELO$ with a
beam splitter (BS) results in a fringe pattern recorded with
a CCD camera (PCO Pixelfly QE, $1392 \times 1024$ square pixels of
6.7 $\mu \rm m$, frame rate $\omegaS = 12 \, \rm Hz$). Proper
frequency detuning $\Delta \omega = \omegaTWO - \omegaONE$, and
angular tilt $\theta$ enable accurate and sensitive
phase-shifting off-axis holography \cite{AtlanGross2007,
GrossAtlan2007}.

The first  samples observed are thin layers of gold beads (diameter
$d$ = 50 nm to 200 nm) immobilized in a polyvinyl alcohol (PVA)
matrix spread by spin coating onto a glass slide. These slides are
set onto a prism used to guide the illumination field $\EI$ and
provoke TIR at the PVA-air interface. Microscope immersion oil
between the prism and the slide enables refractive index matching.
The evanescent wave locally frustrated by the beads is turned into a
propagative scattered field $\EO$ collected by the MO. To make $\EO$
and $\ELO$ in phase opposition from one frame to the next, two-phase
detection ($\Delta \omega=\omega_S/2$) is performed. Exposure time
is $\TE = 50 \, \rm ms$, and holograms are obtained by making the
difference of consecutive
frames.
The complex field $E(x,y,z)$, in the beads plane $z$, is
reconstructed by numerical Fresnel transform \cite{AtlanGross2007}.

\begin{figure}[]
\centering
\includegraphics[width = 7.5 cm]{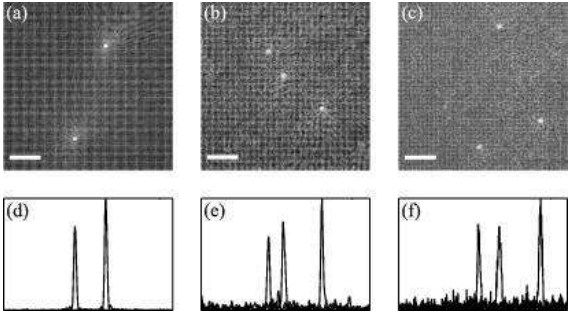}
\caption{Squared amplitude holograms $|E|^2$ in logarithmic
arbitrary units. 200 nm beads (a), 100 nm
beads (b), and 50 nm beads (c). Scale bar is 10 $\mu \rm m$.
Horizontal profile traces of $|E|^2$ at the beads positions,
averaged over 4 pixels (d, e, f), in linear arbitrary units.
}\label{fig_070227_200_100_50}
\end{figure}

\begin{figure}[]
\centering
\includegraphics[width = 8.5cm]{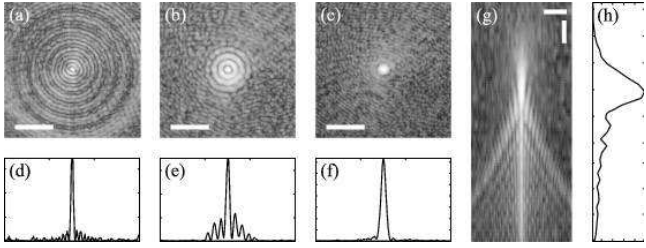}
\caption{$|E|^2$ reconstructed images of a 200 nm gold bead with
experimental setup of fig.\ref{fig_setup}(b) at several axial
positions : $z = 22\, \mu \rm m$ (a) $z = 17\, \mu \rm m$ (b) $z =
0\, \mu \rm m$ (c), displayed in logarithmic arbitrary units. Scale bar is 5 $\mu \rm m$. Transverse plane
profile traces of $|E|^2$ at the beads positions, averaged over 3
pixels ((d) to (f)), in linear arbitrary units. Axial plane
distribution (g) and linear scale profile averaged over 10 pixels
in the lateral direction (h).} \label{fig_zpsf}
\end{figure}

Fig.\ref{fig_070227_200_100_50} shows the bead images (a,b,c), and
the corresponding intensity profiles  along $x$ (d,e,f). The  images
are obtained by averaging  ($|\EO(x,y,z)|^2$) over 4 sequences of 2
images. The beads diameters are 200 (a,d), 100 (b,e) and 50 nm
(c,f). Under 50 nm the coherent parasitic light scattered by dust or
surface roughness prevents the beads to be distinguished.
Fig.\ref{fig_zpsf} shows, for 200 nm beads, the $x,y$ images (a,b,c)
and the $x$ profiles (d,e,f), at $z$ = 0, 17, and 22 $\mu \rm m$
relative reconstruction distances. It also shows the $x,z$ image (g)
and the $z$ profile (h), in the $z$ = 0 to 47 $\mu \rm m$ range.
Because the bead emitters are close to a plane dielectric interface
\cite{lukosz15lem}, one observes an asymmetric signal in the $z$
direction of $\sim 7 \, \mu \rm m$ width at half maximum with rings
in the $x,y$ plane. Similar rings have been observed with quantum
dot emitters \cite{speidel2003tdt, patra2005diq}. Quantitative study
of this effect \cite{lukosz15lem, mertz2000raf}, is out of the scope
of this letter.

\begin{figure}[]
\centering
\includegraphics[width = 8.5 cm]{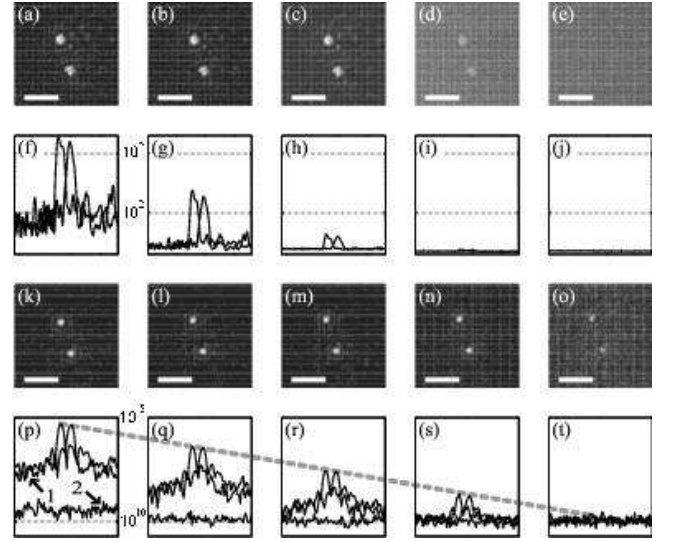}
\caption{Intensity images calculated from
$4\times 2$ camera frames data of a pair of 200 nm particles. (a) to (e) :
direct image (LO beam is off). (k) to (o) : holography (LO beam is
on). Scale bar is 5 $\mu \rm m$. Illumination field power
attenuation ranges from $10^0$ to $10^{4}$. Bead profiles without (f) to (j), and with LO beam (p) to (t). The vertical
axis range (A.U.) is the same for (f) to (j), and
(p) to (t). }\label{fig_070305all_attenuation_crop}
\end{figure}

To illustrate the sensitivity of our holographic setup, we have
displayed on Fig. \ref{fig_070305all_attenuation_crop} images ((a,
..., e) and (k, ..., o)) and profiles ((f, ..., j) and (p, ..., t))
of 200 nm beads at different illumination intensities. The
illumination beam ($|\EI|^2$) is attenuated by a factor $10^0$
(a,f,k,p), $10^1$ (b,g,l,q) ..., $10^{4}$ (c,j,o,t). The LO beam is
either off (direct imaging : (a, ..., j)), or on (holographic regime
: (k, ..., t)). The beads are imaged  with  much better
sensitivity with the LO beam, since the bead signal remains visible
by reducing the illumination power over 4 orders of magnitude.
Moreover, the signal is proportional to illumination (dashed line
throughout Fig.\ref{fig_070305all_attenuation_crop}(p) to (t)). We
must note that, at high illumination level (p), the bead detection
dynamic range is limited by the parasitic light background (arrow 1)
which is several orders of magnitude larger than the holographic
detection noise floor (arrow 2), obtained in a cut from the
off-axis quietest region of the reconstructed
image, away from the beads region.  
The noise floor (about $10^{10}$ A.U.), which is related to shot
noise of the LO beam, gives an absolute calibration of the bead peak
($5 \times 10^{14}$ A.U. on (k)), since it corresponds to 1 photo electron
(e) per pixel \cite{GrossAtlan2007}. Without attenuation (k), the
bead peak area is about $20 \times 20$ pixels (see (p)); it
corresponds thus to  $\sim 2 \times 10^7$ e. One watt of laser yields $\sim 10^{19}$ photons per second. With $10$ mW and $2 \TE = 100 $ ms (2 frames to make an
hologram), we get $10^{16}$ photons. The Mie scattering cross
section of 200 nm gold particles (refractive index : $n = 0.39 -2.38 j$ at
532 nm; with $j^2=-1$) \cite{mie} 
is $0.1 \mu \textrm{m}^2$. The total illumination area is $\sim 1 \textrm{mm}^2$.
Thus we get $10^9$ scattered photons per bead. Since the bead is located at
an air-glass interface, most of the light ($85 \%$) is scattered
within the glass \cite{brokmann2004het}, and $1.5 \times 10^8$ photons are
scattered forward within a $2\pi$ solid angle. The collection
solid angle of a NA = 0.9 objective is $\sim \pi/4$. We then get $4 \times 10^7 $ photons on the CCD yielding an expected signal of $2 \times 10^7$ e
($50\%$ quantum efficiency at 532 nm), in agreement with the
experiment.

To assess imaging performances in experimental conditions compatible with biological microscopy, we have performed axial sectioning of
dynamic beads; we have imaged a suspension of 200 nm beads in
brownian motion in water (see Fig.\ref{fig_setup}(c)). A $\sim 30 \,
\mu \rm m$-thick layer sample is realized within a slide / parafilm
(TM) / coverslip stack. The parafilm layer was heated until melting
to serve as waterproof spacer of $\sim$ 30 microns thickness. In
this configuration, the top interface (coverslip-air) is the one
where TIR occurs. The part of the illumination field $\EI$ not
diffracted by the beads undergoes TIR while the scattered field
$\EO$ passes through the MO and is used for imaging. Since the
exposure time is short : $\TE=1$ ms, the LO power is
increased to fill the CCD dynamic range. Over a time $\tau$, brownian
particles travel by a distance $r(\tau)= (6 \DB \tau)^{1/2}$, where
$\DB = 2.1 \times 10^{-12} \, \rm
m^2 . s^{-1}$ is the diffusion coefficient of a 200 nm particle in water. During one exposure, this travel is smaller than the particle size: $r (1~\textrm{ms}) \sim 110 \, \rm
nm$. 200 nm beads thus appear to be quasi immobile, but travel from one image to
the next by $r(80~\textrm{ms}) \sim 1 \, \mu \rm m $. This property
is used for signal demodulation. The hologram is obtained by
subtracting from the current recorded frame the time average of the
10 next consecutive frames. In this way the holographic information
is recorded in one exposure $\TE = 1$ ms. To localize smaller beads,
the exposure time might be shortened accordingly.

\begin{figure}[]
\centering
\includegraphics[width = 8.5 cm]{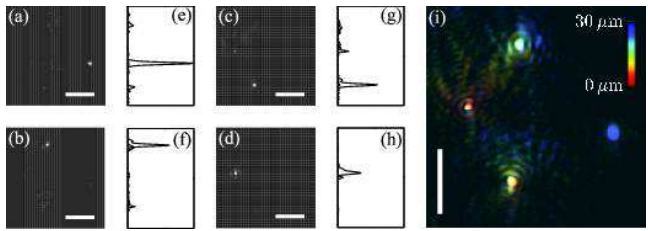}
\caption{Axial exploration of a 200 nm gold beads suspension in
water. $|\EO|^2$ images reconstructed 10 $\mu \rm m$ apart (a) to
(d). Linear scale profiles averaged over 4 pixels (e) to (h).
Composite image (i). Scale bar is 5 $\mu \rm
m$.}\label{fig_070511_volume3D_flat}
\end{figure}

Fig.\ref{fig_070511_volume3D_flat} shows a series of images
(a,b,c,d) and profiles (e,g,f,h) calculated from a
single hologram. Images (a) to (d) and profiles (c) to (h) are
calculated at axial positions $z$ (a,e), $z+10 \,\mu \textrm{m}$ (b,f),... $z + 30
\,\mu \textrm{m}$ (d,h). A composite image (i) is made from reconstructed intensity holograms at axial positions ($z$) coded in color in the 30 $\mu$m range. The attached multimedia file shows the video rate motion of the particles in this composite image.

In this letter we have performed full field imaging of gold nano
particles in 3D at video-rate with exposure times as short as
$1~\textrm{ms}$. Our holographic technique, which benefits from
heterodyne gain, exhibits optimal detection sensitivity, and, since
the signal can be easily calibrated, we have verified its agreement with the expected signal.
In the future, better axial resolution might be obtained
by using the phase of the holographic data. Low-coherence light
sources might be used to reduce the background light component and
improve depth-sectioning. The background might also be reduced by a selective modulation (e.g. spatial or photothermal) of the particles.

The authors acknowledge support from the French Agence Nationale de la Recherche
(ANR) and the Centre de comp\'etence NanoSciences \^Ile de France
(C'nano IdF).

Contact author : atlan@lkb.ens.fr


\bibliographystyle{apsrev}
\bibliographystyle{unsrt}

\end{document}